# Statistics dependent spectral properties of random arrays of particles

**ROMIL AUDHKHASI[1], MAKSYM ZHELYEZNYAKOV[1], STEVEN BRUNTON[2] AND ARKA MAJUMDAR[1,3,*]**

[1]*Department of Electrical and Computer Engineering, University of Washington, Seattle, Washington 98195, United States*
[2]*Department of Mechanical Engineering, University of Washington, Seattle, Washington 98195, United States*
[3]*Department of Physics, University of Washington, Seattle, Washington 98195, United States*
*[*]arka@uw.edu*

**Abstract:** The ability to tailor the spectral response of photonic devices is paramount to the advancement of a broad range of applications. The vast design space offered by disordered optical media provide enhanced functionality for spectral tailoring, although it is challenging to map the spectral properties of such complex systems to their structural attributes. In this work, we investigate correlations between the statistics underlying the structure of random arrays of particles and their spectral properties. We consider 1D and 2D arrays of dielectric nanorods suspended in vacuum and numerically study their optical scattering properties in the 300 – 800 nm wavelength range. We show that the scattering cross section of a random particle array is primarily governed by its configuration statistics and is independent of its exact instantiation or the number of its constituent particles. We further exploit the strong correlations between the statistics and spectral properties of random particle arrays to predict their spectral response. By using a semi-analytical nearest neighbor coupling model, we produce accurate qualitative estimates of the spectral responses of both one and two-dimensional random arrays for different configuration statistics. The results presented in this manuscript will open new avenues for optimizing large-scale random systems to achieve enhanced optical functionalities for a wide variety of applications.

## 1. Introduction

Nature is abundant with randomness and disorder, and random media play a critical role in sustaining the environment around us [1, 2]. In particular, the interaction of light with such complex systems gives rise to various interesting physical phenomena. For instance, the brilliant colors found in butterfly wings are the result of light scattering from intricate arrangements of scales that generate structural coloration over a wide range of viewing angles [3-6]. On the other hand, white beetles use scales with disordered photonic structures to achieve exceptional whiteness for camouflage [7].

Historically, ordered and periodic photonic structures have received considerable attention from the scientific community [8] and randomness and disorder have often been viewed as undesirable. Recently, owing to the wide range of functionalities exhibited by disordered systems found in nature, researchers have started investigating their optical properties. By judiciously engineering the nature and extent of randomness in such systems, several studies have proposed disordered photonic devices for applications in spectroscopy [9, 10], imaging and lasing [11].

One of the most important attributes of any photonic device is its spectral response. The ability to tailor the spectral response of a given photonic device is of great significance for a broad range of applications from sensing [12-14] and spectroscopy to thermal management [15-

18] and energy harvesting [19, 20]. A majority of the literature on photonic devices with tailored spectral responses has focused on highly ordered systems supporting either localized [21, 22] or delocalized [23-25] electromagnetic (EM) modes. Owing to the large number of degrees of freedom offered by disordered media, it is reasonable to expect such systems to provide greater flexibility for spectral tailoring compared to their ordered counterparts. However, the vast design space associated with random photonic devices also makes it challenging to map their structural degrees of freedom to their spectral properties. Therefore, it is of immense importance to determine a tractable set of parameters that can be used to characterize and predict the spectral response of such complex systems.

In this work, we investigate the correlation between the statistics governing the structural attributes of a random particle array and its spectral properties. We consider 1D and 2D arrays of square dielectric nanorods suspended in vacuum. The system is characterized by the side lengths of the nanorods and the distances between them. We begin by showing that the spectral response of a system having a predefined set of side lengths and distances can be modulated by simply changing the statistics governing the values of these parameters. Moreover, for predefined statistics, the normalized spectral response of the system is independent of the number of its constituent particles. Finally, we explore how knowledge about the statistics governing a given random system can be used to predict its spectral response.

We believe that the results presented in this manuscript will provide new avenues for understanding optical phenomena in random photonic devices. The ability to characterize, predict and tailor the spectral response of large scale random media will be useful for several applications in spectroscopy and imaging.

## 2. Results and discussion

### 2.1 Correlation between configurational statistics and spectral properties of a random particle array

We begin by investigating the correlation between the statistics governing the structural attributes of a given random system and its spectral response. We consider a random system comprised of a one-dimensional (1D) array of 100 square nanorods made of $Si_3N_4$ (Fig. 1(a)). For simplicity, we assume that the side length of each nanorod in the array (denoted by $L$) can take two possible values: $L_1 = 0.5$ μm with probability $p_{L1}$ and $L_2 = 1$ μm with probability $p_{L2}$. Similarly, the edge-to-edge distance between any two adjacent nanorods (denoted by $d$) can take two possible values: $d_1 = 0.1$ μm with probability $p_{d1}$ and $d_2 = 0.4$ μm with probability $p_{d2}$. The nanorods are assumed to be infinitely long in the direction perpendicular to the $x-y$ plane (see Fig. 1(a) for coordinate system). We consider four nanorod arrays with different $L$ and $d$ statistics represented by the probability density functions shown in the left and middle panels of Figs. 1(b) through (d). The corresponding scattering cross sections, $\sigma_S$ in the 300 – 800 nm wavelength range are displayed in the right panels of Figs. 1(b) through (d). In each case, the solid black line represents the mean $\sigma_S$ for 10 instantiations of the array while the blue band represents the range (difference between the highest and lowest value for each wavelength). In the context of this work, an instantiation of the array corresponds to a specification of the sizes and locations of all the nanorods.

The scattering cross sections are calculated using Lumerical FDTD. The optical constants of $Si_3N_4$ are taken from Ref. [26]. We use Perfectly Matched Layer (PML) boundary conditions along the x- and y- directions. The structure is illuminated by an *x*-polarized Total Field Scattered Field (TFSF) source incident normally along the y-direction. The scattering cross section is recorded as a function of wavelength using a two-dimensional (2D) power flux monitor. The coordinate system along with the direction of the incident plane wave ($k$) and its electric field ($E$) are displayed in Fig. 1(a). Here we present the scattering cross sections normalized to the total lateral span of the array, defined as the distance between the left edge of the first and the right edge of the last nanorod.

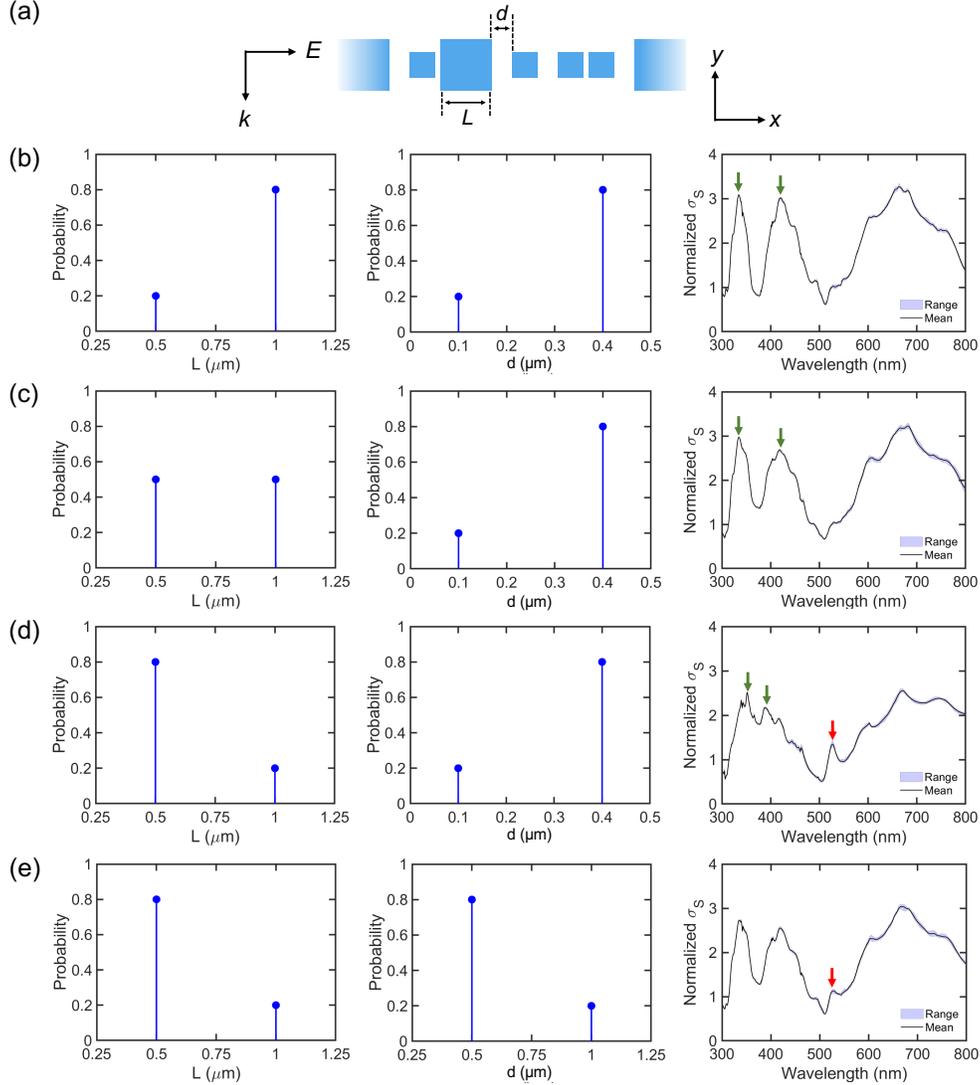

Fig 1. (a) Schematic showing a 1D random array of 100 square $Si_3N_4$ nanorods in which $L$ can take a value of $L_1 = 0.5$ μm or $L_2 = 1$ μm and $d$ can take a value of $d_1 = 0.1$ μm or $d_2 = 0.4$ μm. (b) – (e) (left panel) Probability density function for $L$ statistics, (right panel) probability density function for $d$ statistics; and (right panel) normalized scattering cross section at normal incidence for four different arrays. In each case, the solid black line shows the mean scattering cross section of 10 different instantiations of the array while the blue band shows the range. The relevant spectral features referred to in the text are highlighted by colored arrows.

From Figs. 1(b) – (e), it can be inferred that all instantiations of the random array with fixed $L$ and $d$ statistics have a similar spectral response. The data presented in Figs. 1(b), (c) and (d) corresponds to arrays with the same $d$ statistics ($p_{d1} = 0.2$ and $p_{d2} = 0.8$) but different $L$ statistics. We observe that a change in the nanorod side length probability values alone causes a significant change in the spectral response of the array. For instance, the broad spectral features observed at 350 nm and 420 nm (indicated by the dark green arrows in Figs. 1(b) – (d)) for $p_{L1} = 0.2$, $p_{L2} = 0.8$ undergo a reduction in amplitude for $p_{L1} = 0.5$, $p_{L2} = 0.5$ and an increase in spectral overlap for $p_{L1} = 0.8$, $p_{L2} = 0.2$. Similarly, we observe the emergence of a spectral feature at a wavelength of 530 nm (red arrow in Fig. 1(d)) in going from Fig. 1(b) to (d). Figure 1(e) shows the spectral response of a random array with the same $L$ statistics as Fig. 1(d) but different $d$ statistics ($p_{d1}$ = 0.8 and $p_{d2}$ = 0.2). By modifying the $d$ statistics, we observe a reduction in the amplitude of the spectral feature at 530 nm in Fig. 1(e) relative to Fig. 1(d). From the data presented in Figs. 1(b) – (e), we conclude that the spectral response of a random array of nanorods with predefined length and edge-to-edge separation values is governed solely by its configuration statistics and not by its exact instantiation. On the other hand, we note that the spatial distribution of the electromagnetic field is highly dependent on the exact instantiation of the random array, and as such has no specific dependence on the underlying statistics.

Next, we explore the effect of the number of particles in a random system with predefined statistics on its spectral properties. We consider two random systems: a 1D array of $N$ square nanorods made of $Si_3N_4$ (Fig. 2(a)), and a 2D array of $N$ cubic $Si_3N_4$ nanorods (Fig. 2(b)). For both systems, we choose $L_1$ = 0.1 µm and $L_2$ = 0.5 µm with probabilities $p_{L1}$ = 0.2 and $p_{L2}$ = 0.8. The nanorods in the 1D system are assumed to be infinitely long in the direction perpendicular to the plane of the figure. The 2D system is assumed to be pseudo-random with the center-to-center separation between neighboring nanorods fixed at $d$ = 0.7 µm. Given the values of the nanorod sizes and their respective probabilities, this gives two possible values for the edge-to-edge distances for the 2D system: $d_1$ = 0.2 with probability $p_{d1}$ = 0.6 and $d_2$ = 0.4 with probability $p_{d2}$ = 0.4. We use the same values of the edge-to-edge separations and their probabilities to define our 1D system.

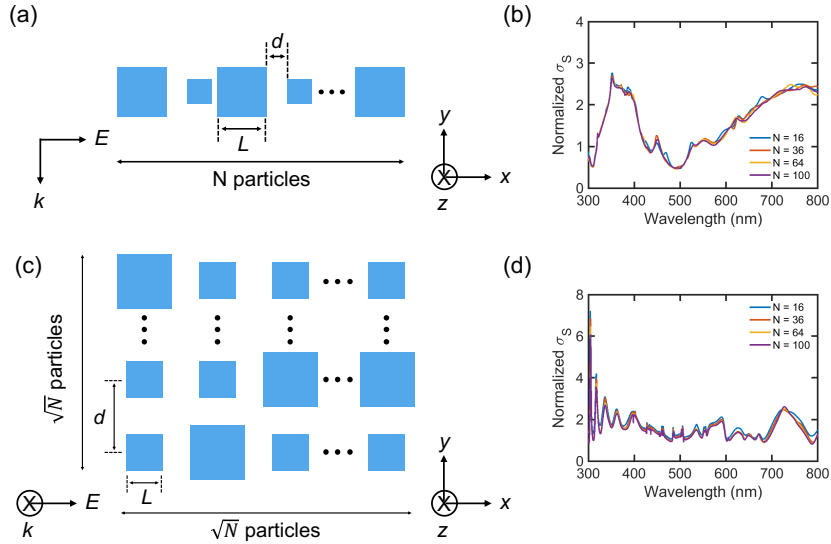

**Figure 2**: (a) Schematic showing a 1D array of $N$ square $Si_3N_4$ nanorods. Here, $L_1$ = 0.1 µm, $L_2$ = 0.5 µm, $d_1$ = 0.2 µm, $d_2$ = 0.4 µm, $p_{L1}$ = 0.2, $p_{L2}$ = 0.8, $p_{d1}$ = 0.6 and $p_{d2}$ = 0.4. (b) Normalized scattering cross section of the array at normal incidence for $N$ = 16, 36, 64 and 100. (c) Schematic showing a 2D array of cubic $N$ $Si_3N_4$ nanorods with the same $L$ values and statistics as the system in (a) and fixed $d$ = 0.7 µm. (d) Normalized scattering cross section of the array at normal incidence for $N$ = 16, 36, 64 and 100.

Figures 2(b) and 2(d) present the scattering cross sections ($\sigma_S$) of the 1D and the 2D system respectively at normal incidence in the 300 – 800 nm wavelength range for different values of the number of particles $N$. For the 1D system, we normalize $\sigma_S$ to the lateral span of the system while for the 2D case, we normalize it to the area of the smallest rectangle circumscribing the system. For calculating the scattering cross sections of the 2D array in FDTD simulations, we illuminate it with an x-polarized TFSF source incident along the z direction (see Fig. 2(c) for the coordinate system). We observe that for both systems, the normalized scattering cross sections are independent of the number of particles. The unnormalized $\sigma_S$ (not shown here) reduces with an increase in the number of particles but retains the same spectral profile. Additionally, we observe that for fixed $N$ and configuration statistics, the 2D system has more spectral uniformity than the 1D system. While the 1D system is primarily associated with broad spectral features, the spectrum of the 2D system has both narrow and broad linewidth peaks. These observations suggest that the spectral properties of random arrays of particles are independent of the number of their constituent particles and are instead dictated by their statistics. Moreover, for fixed values of the structural attributes and their statistics, the spectral properties can be modulated by either arranging the particles in a 1D or a 2D array.

In the next subsection, we explore if the strong correlation between the optical properties of random arrays of particles and their statistics can be utilized for spectral prediction in such complex systems.

*2.2 Using statistics to predict the spectral response of random particle arrays*

Let us consider a situation in which starting with a random system with predefined nanorod sizes and inter-nanorod distances, we wish to optimize the statistics to achieve a desired spectral response. In this case, it is advantageous to have a way of predicting the spectral response of the system from its configuration statistics. As a first approximation, one can write the spectral response of the full system as an average of the spectral responses of each type of particle in the system weighted by the nanorod size statistics. However, such an approximation considers all the particles in the system to be isolated and neglects the effect of coupling.

In general, the spectral response of each particle in the array is influenced by the scattering properties of all the other particles in the array. Here, we propose a spectral prediction approach based on the nearest neighbor coupling approximation. In this case, the spectral response of a given particle in the array is only influenced by its nearest neighbors. We show that the spectral response of the full array can be written as a weighted sum of the spectral responses of a small number of *particle clusters*.

We first illustrate the utility of this approach for predicting the spectral response of a 1D random array of $N$ square $Si_3N_4$ nanorods (Fig. 3(a)). The nanorods have a side length $L_i$ with probability $p_{Li}$ and an edge-to-edge separation $d_i$ with probability $p_{di}$ ($i = 1, 2$). As an example, we choose $N = 100$, $L_1 = 0.4$ μm, $L_2 = 0.5$ μm, $d_1 = 0.1$ μm and $d_2 = 1$ μm. For this system, we define the particle cluster as a pair of nanorods with side lengths $L_a$ and $L_b$ separated by a distance $d$ (indicated by the red box in Fig. 3(a)). Since each of $L_a$, $L_b$ and $d$ can take two possible values, there are eight possible particle clusters. We determine their scattering cross sections from FDTD simulations. The spectral response of the full array of $N$ nanorods can then be written in terms of the scattering cross sections of the particle clusters as:

$$\sigma_S = \frac{N}{N_{counted}} \sum_{i=1}^{8} \left( p_{cluster,i} N_{clusters} \right) \sigma_{S\,cluster,i}$$

*( 1 )*

Here $N_{clusters}$ denotes the total number of particle clusters in the system while $p_{cluster,I}$ and $\sigma_{S\,cluster,i}$ denote the fraction of all clusters that are of type $i$ ($i \in [1, 8]$) and their corresponding scattering cross section, respectively. We consider the system to be composed of contiguous nanorod pairs and hence $N_{clusters} = N - 1$. The assumption of contiguous particle clusters results in an overcounting of the number of particles in the system. To account for this, we readjust the predicted scattering cross section by multiplying it with the factor $N/N_{counted}$. As all particles except the first and the last are counted twice, $N_{counted} = 2(N-1)$.

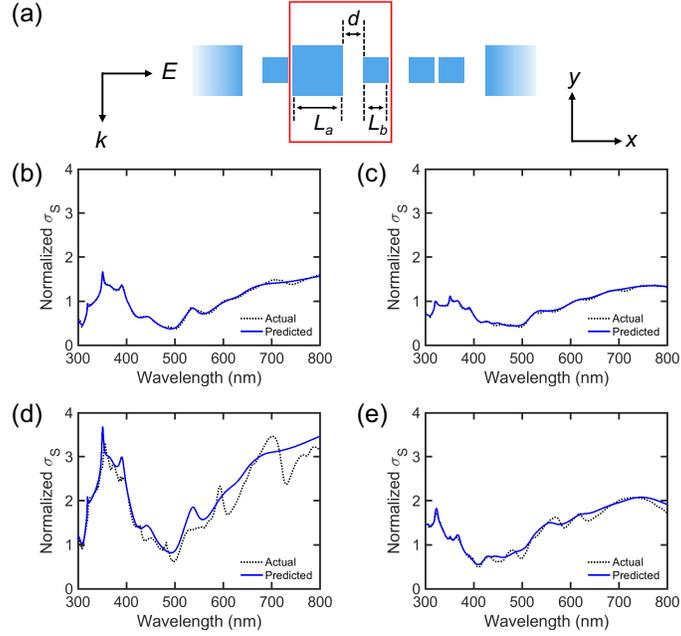

Fig 3. (a) Schematic showing a 1D random array of 100 Si$_3$N$_4$ nanorods with a single particle cluster indicated by the red box. Each of $L_a$ and $L_b$ can have two possible values: $L_1 = 0.4$ μm and $L_2 = 0.5$ μm, while $d$ can either have a value of $d_1 = 0.1$ μm or $d_2 = 1$ μm. (b) – (e) Predicted and actual scattering cross section of the array at normal incidence for: (b) $p_{L1} = 0.2$, $p_{L2} = 0.8$, $p_{d1} = 0.2$, $p_{d2} = 0.8$; (c) $p_{L1} = 0.5$, $p_{L2} = 0.5$, $p_{d1} = 0.5$, $p_{d2} = 0.5$; (d) $p_{L1} = 0.2$, $p_{L2} = 0.8$, $p_{d1} = 1$, $p_{d2} = 0$; and (e) $p_{L1} = 0.8$, $p_{L2} = 0.2$, $p_{d1} = 0.8$, $p_{d2} = 0.2$.

Figures 3(b) – (d) present the actual and predicted scattering cross sections of four random arrays with the same values of $N$, $L_1$, $L_2$, $d_1$ and $d_2$ but different $L$ and $d$ statistics. In each case, the cross sections are normalized to the lateral spans of the arrays. We observe that while the predictions match well with the actual spectral responses in Figs. 3(b) and (c), significant deviations occur in Figs. 3(d) and (e). To quantify these deviations, we define a percentage relative error between the actual and predicted spectra, as given by the following equation:

$$\Delta \sigma_S = 100 \; x \; \frac{\int_{\lambda_1}^{\lambda_2} |\sigma_{S,actual}(\lambda) - \sigma_{S,pred}(\lambda)| d\lambda}{\int_{\lambda_1}^{\lambda_2} \sigma_{S,actual}(\lambda) d\lambda}$$

(2)

For the arrays corresponding to Figs. 3(b) and (c), $\Delta \sigma_S = 2.5\%$ and 1.9%, while for Figs. 3(d) and (e), $\Delta \sigma_S = 11.2\%$ and 4.5%, respectively. The larger errors in our model predictions for Figs. 3(d) and (e) can be understood by noting that the random arrays corresponding to these two cases have small values of mean inter-particle distances: 0.1 and 0.28 μm, respectively. At such small distances, long range coupling exists between the nanorods constituting the arrays. This causes the actual spectral response to differ considerably from that obtained from the nearest neighbor coupling assumption. Despite the deviations, our model provides a good qualitative estimate of the spectral responses of the arrays in Figs. 3(d) and (e).

As a second example, we discuss spectral prediction for a 2D pseudo-random array of $N = 100$ cubic Si$_3$N$_4$ nanorods with sizes $L_1 = 0.1$ μm and $L_2 = 0.5$ μm and fixed center-to-center distance $d = 0.7$ μm. For this system, we choose the particle cluster to be a triplet of nanorods with side lengths $L_a$, $L_b$ and $L_c$ (indicated by the red outline in Fig. 4(a)). Since each nanorod in the triplet can have two possible values for the side length, there are a total of eight possible particle clusters. We determine the spectral responses of these clusters from FDTD simulations. The scattering cross section for the full 2D array assuming nearest neighbor coupling is given by Eq. 1 with $N_{clusters} = 2(N-1)^2$ and $N_{counted} = 5(\sqrt{N}-2)^2 + 8(\sqrt{N}-2) + N + 2$.

Figures 4(b) through (e) show the predicted and actual scattering cross sections of four arrays with the same $(p_{L1}, p_{L2}) = (0.5, 0.5), (0.8, 0.2), (0, 1)$ and $(0.2, 0.8)$ respectively. In each case, the scattering cross section is normalized to the area of the smallest rectangle circumscribing the array. We observe that in all four cases, our nearest neighbor coupling model provides a good qualitative estimate of the spectral response of the full 2D array. However, the difference between the predicted and actual scattering cross sections is much lower for the systems in Figs. 4(b) and (c) than for those in Figs. 4(d) and (e). The relative error $\Delta\sigma_S$ between the actual and predicted spectra in Figs. 4(b) and (c) is 14.8% and 12.3% while for Figs. 4(d) and (e), $\Delta\sigma_S = 22.9\%$ and 19%, respectively. This difference in error can be understood by noting that for the systems in Figs. 4(d) and 4(e), majority of the particles are identical and have a small edge-to-edge distance of 0.2 µm. At such small separations, the coupling of each particle in the array to particles other than its immediate neighbors must be accounted for in order to accurately predict the spectral response of the system.

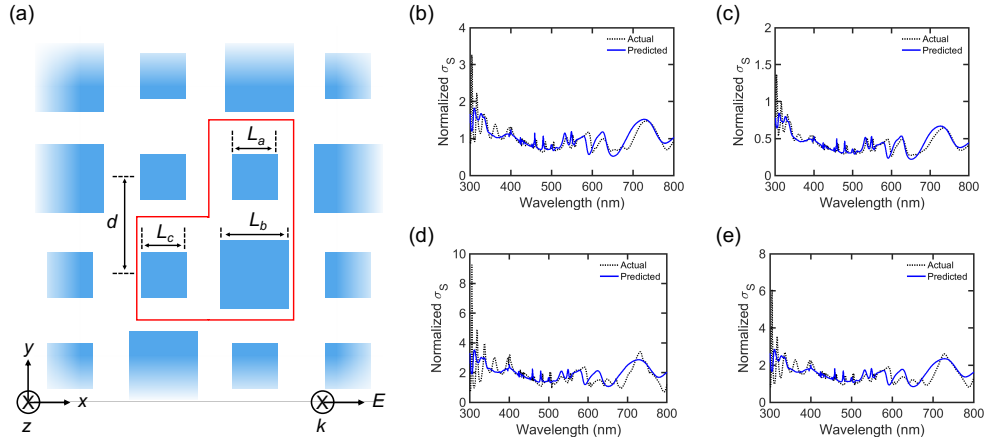

Fig 4. (a) Schematic showing a 2D pseudo – random array of 100 cubic $Si_3N_4$ nanorods with a single particle cluster indicated by the red outline. Each of $L_a$, $L_b$ and $L_c$ can have two possible values: $L_1 = 0.1$ µm and $L_2 = 0.5$ µm while $d$ €fixed at 0.7 µm. (b) – (e) Predicted and actual scattering cross section of the array at normal incidence for: (b) $p_{L1} = 0.5$, $p_{L2} = 0.5$; (c) $p_{L1} = 0.8$, $p_{L2} = 0.2$; (d) $p_{L1} = 0$, $p_{L2} = 1$ an $p_{L1} = 0.2$, $p_{L2} = 0.8$.

## 3. Conclusion

In this work, we investigated the correlation between the statistics governing the structural attributes of a random system and its spectral properties. For simplicity, we considered random systems comprised of arrays of $Si_3N_4$ nanorods suspended in vacuum and studied their scattering cross sections in the 300 – 800 nm wavelength range. We began by showing that the spectral response of a random system with predefined values of structural attributes is governed solely by its configuration statistics and is independent of its exact instantiation or the number of constituent particles. As a consequence of this, for a given random system with $N$ particles, one can always identify an equivalent random system with a smaller number of particles that has the same spectral response. This provides a route to determining the spectral properties of large scale random systems for which full wave simulations may not be feasible.

In the latter part of the manuscript, we exploited the strong correlations between the optical properties of random particle arrays and their statistics for spectral prediction. By using a semi-analytical nearest neighbor coupling model, we were able to provide good qualitative estimates of the spectral responses of both 1D and 2D random systems for different configuration statistics. However, for random systems with strong inter-particle coupling, the predictions of our model showed substantial deviations from the simulated spectral responses. Future work may incorporate the effect of configuration statistics on inter-particle coupling into the prediction model to obtain more accurate quantitative estimates of the spectral response. Furthermore, our model can potentially be generalized to three – dimensional systems by expressing the predicted spectrum using non-linear functions of the particle cluster spectral responses, instead of a simple linear superposition.

We believe that the results presented in this manuscript provide new insights into the spectral properties of random media. The strong correlations between the optical properties of such complex systems and their statistics create new avenues for predicting and optimizing their topology to achieve enhanced optical functionalities for a broad range of applications.

## Funding

This work was supported by the DARPA Coded Visibility STTR program.

## Disclosures

The authors declare no conflicts of interest.

## Data Availability

Data underlying the results presented in this paper are not publicly available at this time but may be obtained from the authors upon reasonable request.